\documentclass[11pt]{elsart}
\usepackage{amssymb}
\usepackage{amsfonts}
\usepackage{slashbox}
\usepackage{graphicx}
\usepackage{amsmath}
\usepackage{mathrsfs} 
\usepackage{txfonts}

\begin{document}
\begin{frontmatter}
\title{Quantum Bayesian implementation}

\begin{abstract}
Bayesian implementation concerns decision making problems when
agents have incomplete information.  A recent work [Wu, Quantum
mechanism helps agents combat ``bad'' social choice rules.
\emph{Intl. J. of Quantum Information} 9 (2011) 615-623] generalized
the implementation theory with complete information to a quantum
domain. In this paper, we propose a quantum Bayesian mechanism and
an algorithmic Bayesian mechanism, which amend the traditional
results for Bayesian implementation.
\end{abstract}
\end{frontmatter}

\section{Introduction}
Mechanism design is an important branch of economics. Compared with
game theory, it concerns a reverse question: given some desirable
outcomes, can we design a game that produces them? Nash
implementation and Bayesian implementation are two key topics of the
mechanism design theory. The former assumes complete information
among the agents, whereas the latter concerns incomplete
information. Maskin \cite{Maskin1999} provided an almost complete
characterization of social choice rules that are Nash implementable
when the number of agents is at least three. Postlewaite and
Schmeidler \cite{PS1986}, Palfrey and Srivastava \cite{PS1989}, and
Jackson \cite{Jackson1991} together constructed a framework for
Bayesian implementation.

In 2011, Wu \cite{qmd2011} claimed that the sufficient conditions
for Nash implementation shall be amended by virtue of a quantum
mechanism. Furthermore, this amendment holds in the macro world by
virtue of an algorithmic mechanism \cite{sim2011}. Given these
accomplishments in the field of Nash implementation, this paper aims
to investigate what will happen if the quantum mechanism is applied
to Bayesian implementation.

The rest of this paper is organized as follows: Section 2 recalls
preliminaries of Bayesian implementation given by Serrano
\cite{Serrano2004}. In Section 3, a novel condition, multi-Bayesian
monotonicity, is defined. Section 4 and 5 are the main parts of this
paper, in which we will propose quantum and algorithmic Bayesian
mechanisms respectively. The last section draws the conclusions.

\section{Preliminaries}
Let $N=\{1,\cdots,n\}$ be a finite set of \emph{agents} with $n\geq
2$, $A=\{a_{1},\cdots,a_{k}\}$ be a finite set of social
\emph{outcomes}. Let $T_{i}$ be the finite set of agent $i$'s types,
and the \emph{private information} possessed by agent $i$ is denoted
as $t_{i}\in T_{i}$. We refer to a profile of types
$t=(t_{1},\cdots,t_{n})$ as a \emph{state}. Consider environments in
which the state $t=(t_{1},\cdots,t_{n})$ is not common knowledge
among the $n$ agents. We denote by $T$ the set of states compatible
with an environment, i.e., a set of states that is common knowledge
among the agents. Let $T=\prod_{i\in N} T_{i}$. Each agent $i\in N$
knows his type $t_{i}\in T_{i}$, but not necessarily the types of
the others. We will use the notation $t_{-i}$ to denote
$(t_{j})_{j\neq i}$. Similarly, $T_{-i}=\prod_{j\neq i} T_{j}$.

Each agent has a \emph{prior belief}, probability distribution,
$q_{i}$ defined on $T$. We make an assumption of nonredundant types:
for every $i\in N$ and $t_{i}\in T_{i}$, there exists $t_{-i}\in
T_{-i}$ such that $q_{i}(t)>0$. For each $i\in N$ and $t_{i}\in
T_{i}$, the conditional probability of $t_{-i}\in T_{-i}$, given
$t_{i}$, is the \emph{posterior belief} of type $t_{i}$ and it is
denoted $q_{i}(t_{-i}|t_{i})$. For simplicity, we shall consider
only single-valued rules, i.e., an SCF $f$ is a mapping $f: T\mapsto
A$. Let $\mathcal{F}$ denote the set of SCFs. Given agent $i$'s
state $t_{i}$ and utility function $u_{i}(\cdot, t): \Delta \times
T\mapsto \mathbb{R}$, the \emph{conditional expected utility} of
agent $i$ of type $t_{i}$ corresponding to a social choice function
(SCF) $f: T\mapsto \Delta$ is defined as:
\begin{equation*}
U_{i}(f|t_{i})\equiv\sum\limits_{t'_{-i}\in
T_{-i}}q_{i}(t'_{-i}|t_{i}) u_{i}(f(t'_{-i},t_{i}),(t'_{-i},t_{i})).
\end{equation*}
An \emph{environment with incomplete information} is a list
$E=<N,A,(u_{i}, T_{i},q_{i})_{i\in N}>$. An environment is
\emph{economic} if, as part of the social outcomes, there exists a
private good (e.g., money) over which all agents have a strictly
positive preference. Two SCFs $f$ and $h$ are \emph{equivalent}
($f\approx h$) if $f(t)=h(t)$ for every $t\in T$.

Consider a \emph{mechanism} $\Gamma=((M_{i})_{i\in N}, g)$ imposed
on an incomplete information environment $E$, $g: M\mapsto
\mathcal{F}$. A \emph{Bayesian Nash equilibrium} of $\Gamma$ is a
profile of strategies $\sigma^{*}= (\sigma^{*}_{i})_{i\in N}$ where
$\sigma^{*}_{i}: T_{i}\mapsto M_{i}$ such that for all $i\in N$ and
for all $t_{i}\in T_{i}$,
\begin{equation*}
U_{i}(g(\sigma^{*})|t_{i})\geq
U_{i}(g(\sigma^{*}_{-i},\sigma'_{i})|t_{i}), \quad \forall
\sigma'_{i}: T_{i}\mapsto M_{i}.
\end{equation*}
Denote by $\mathcal{B}(\Gamma)$ the set of Bayesian equilibria of
the mechanism $\Gamma$. Let $g(\mathcal{B}(\Gamma))$ be the
corresponding set of equilibrium outcomes. An SCF $f$ is
\emph{Bayesian implementable} if there exists a mechanism
$\Gamma=((M_{i})_{i\in N},g)$ such that
$g(\mathcal{B}(\Gamma))\approx f$. An SCF $f$ is \emph{incentive
compatible} if truth-telling is a Bayesian equilibrium of the direct
mechanism associated with $f$, i.e., if for every $i\in N$ and for
every $t_{i}\in T_{i}$,
\begin{equation*}
\sum\limits_{t'_{-i}\in T_{-i}}
q_{i}(t'_{-i}|t_{i})u_{i}(f(t'_{-i},t_{i}),(t'_{-i},t_{i}))\geq
\sum\limits_{t'_{-i}\in T_{-i}}
q_{i}(t'_{-i}|t_{i})u_{i}(f(t'_{-i},t'_{i}),(t'_{-i},t_{i})),
\end{equation*}
$\forall t'_{i}\in T_{i}$.

Consider a strategy in a direct mechanism for agent $i$, i.e., a
mapping $\alpha_{i} =(\alpha_{i}(t_{i}))_{t_{i}\in T_{i}}:
T_{i}\mapsto T_{i}$. A \emph{deception} $\alpha=(\alpha_{i})_{i\in
N}$ is a collection of such mappings where at least one differs from
the identity mapping. Given an SCF $f$ and a deception $\alpha$, let
$[f\circ\alpha]$ denote the following SCF:
$[f\circ\alpha](t)=f(\alpha(t))$ for every $t\in T$. For a type
$t_{i}\in T_{i}$, an SCF $f$, and a deception $\alpha$, let
$f_{\alpha_{i}(t_{i})}(t')=f(t'_{-i}, \alpha_{i}(t_{i}))$ for all
$t'\in T$. An SCF $f$ is \emph{Bayesian monotonic} if for any
deception $\alpha$, whenever $f\circ\alpha\napprox f$, there exist
$i\in N$, $t_{i}\in T_{i}$, and an SCF $y$ such that
\begin{equation*}
U_{i}(y\circ\alpha|t_{i})>U_{i}(f\circ\alpha|t_{i}), \quad
\mbox{while }U_{i}(y_{\alpha_{i}(t_{i})}|t'_{i})\leq
U_{i}(f|t'_{i}), \quad\forall t'_{i}\in T_{i}. \quad \mbox{(*)}.
\end{equation*}
In economic environments, the sufficient and necessary conditions
for full Bayesian implementation are incentive compatibility and
Bayesian monotonicity. To facilitate the following discussion, here
we cite the Bayesian mechanism (Page 404, line 4,
\cite{Serrano2004}) as follows: Consider a mechanism
$\Gamma=((M_{i})_{i\in N}, g)$, where $M_{i}=T_{i}\times\mathcal{F}
\times \mathbb{Z}_{+}$, and $\mathbb{Z}_{+}$ is the set of
nonnegative integers. Each agent is asked to report his type
$t_{i}$, an SCF $f_{i}$ and a nonnegative integer $z_{i}$, i.e.,
$m_{i}=(t_{i},f_{i},z_{i})$.
The outcome function $g$ is as follows:\\
(i) If for all $i\in N$, $m_{i}=(t_{i},f,0)$, then $g(m)=f(t)$, where $t=(t_{1},\cdots,t_{n})$.\\
(ii) If for all $j\neq i$, $m_{j}=(t_{j},f,0)$ and
$m_{i}=(t'_{i},y,z_{i})\neq(t'_{i},f,0)$,
we can have two cases:\\
(a) If for all $t_{i}$, $U_{i}(y_{t'_{i}}|t_{i})\leq U_{i}(f|t_{i})$, then $g(m)=y(t'_{i}, t_{-i})$;\\
(b) Otherwise, $g(m)=f(t'_{i}, t_{-i})$.\\
(iii) In all other cases, the total endowment of the economy is
awarded to the agent of smallest index among those who announce the
largest integer.

\section{Multi-Bayesian monotonicity}
\textbf{Definition 1}: An SCF $f$ is \emph{multi-Bayesian monotonic}
if there exist a deception $\alpha$, $f\circ\alpha\napprox f$, and a
set of agents $N^{\alpha}=\{i^{1},i^{2},\cdots\}\subseteq N$, $2\leq
|N^{\alpha}|\leq n$, such that for every $i\in N^{\alpha}$, there
exist $t_{i}\in T_{i}$ and an SCF $y^{i}\in\mathcal{F}$ that
satisfy:
\begin{equation*}
U_{i}(y^{i}\circ\alpha|t_{i})>U_{i}(f\circ\alpha|t_{i}), \quad
\mbox{while }U_{i}(y^{i}_{\alpha_{i}(t_{i})}|t'_{i})\leq
U_{i}(f|t'_{i}), \quad\forall t'_{i}\in T_{i}. \quad \mbox{(**)}.
\end{equation*}
Let $l=|N^{\alpha}|$. Without loss of generality, let these $l$
agents be the last $l$ agents among $n$ agents.

In 1993, Matsushima \cite{Matsushima1993} claimed that Bayesian
monotonicity is a very weak condition when utility functions are
quasi-linear and lotteries are available. Consider an SCF $f$ that
satisfies Bayesian monononicity, if there is a deception $\alpha$
such that its corresponding agent $i$ has another symmetric agent
$j$ (i.e., $i\neq j$, $u_{i}=u_{j}$, $T_{i}=T_{j}$, the prior belief
and posterior belief hold by them are the same), then $f$ is
multi-Bayesian monotonic.

\textbf{Example 1}: Similar to Example 23.B.5 in Ref.
\cite{MWG1995}, here we consider an auction setting with one seller
(i.e., agent 0) and three buyers (i.e., agent 1, 2 and 3). All
buyers' privately observed valuations $t_{i}$ are drawn
independently from the uniform distribution on $[0, 1]$ and this
fact is common knowledge among the agents. Each buyer submits a
sealed bid, $b_{i}\geq 0$ ($i=1, 2, 3$). The sealed bids are
examined and the buyer with the highest bid is declared the winner.
If there is a tie, the winner is chosen randomly. The winning buyer
pays an amount equal to his bid to the seller. The losing buyer does
not pay anything.

Consider the social choice function $f(t)=(x_{0}(t), x_{1}(t),
x_{2}(t), x_{3}(t), p_{0}(t), p_{1}(t), p_{2}(t)$, $p_{3}(t))$, in
which
\begin{align*}
  &x_{1}(t)=1, \quad\mbox{ if } t_{1}\geq t_{2} \mbox{ and }t_{1}\geq t_{3}; \quad =0 \mbox{ otherwise};\\
  &x_{2}(t)=1, \quad\mbox{ if } t_{2}> t_{1} \mbox{ and } t_{2}\geq t_{3}; \quad =0 \mbox{ otherwise};\\
  &x_{3}(t)=1, \quad\mbox{ if } t_{3}> t_{1} \mbox{ and } t_{3}> t_{2}; \quad =0 \mbox{ otherwise};\\
  &x_{0}(t)=0, \quad\mbox{ for all }t;\\
  &p_{1}(t)=-\frac{2}{3}\theta_{1}x_{1}(t);\\
  &p_{2}(t)=-\frac{2}{3}\theta_{2}x_{2}(t);\\
  &p_{3}(t)=-\frac{2}{3}\theta_{3}x_{3}(t);\\
  &p_{0}(t)=-[p_{1}(t) + p_{2}(t) + p_{3}(t)].
\end{align*}
It can be easily checked that the strategies
$b_{i}(t_{i})=\frac{2}{3}t_{i}$ (for $i=1, 2, 3$) constitute a
Bayesian Nash equilibrium of this auction that indirectly yields the
outcomes specified by $f(t)$. Thus, according to Theorem 1
\cite{Jackson1991}, $f$ is incentive compatible and Bayesian
monotonic. Since the three buyers are symmetric, then according to
the definition of multi-Bayesian monotonicity, $f$ is multi-Bayesian
monotonic.

\textbf{Proposition 1}: In economic environments, consider an SCF
$f$ that is incentive compatible and Bayesian monotonic, if $f$ is
multi-Bayesian monotonic, then $f\circ\alpha$ is not Bayesian
implementable by using the traditional Bayesian mechanism, where
$\alpha$ is specified in the definition of
multi-Bayesian monotonicity.\\
\textbf{Proof}: According to Serrano's proof (Page 404, line 33,
\cite{Serrano2004}), all equilibrium strategies fall under rule (i),
i.e., $f$ is unanimously announced and all agents announce the
integer 0. Consider the deception $\alpha$ specified in the
definition of multi-Bayesian monotonicity. At first sight, if every
agent $i\in N$ submits $(\alpha_{i}(t_{i}),f,0)$, then
$f\circ\alpha$ may be generated as the equilibrium outcome by rule
(i). However, For each agent $i\in N^{\alpha}$, he has incentives to
unilaterally deviate from $(\alpha_{i}(t_{i}),f,0)$ to
$(\alpha_{i}(t_{i}),y^{i},0)$ in order to obtain $y^{i}\circ\alpha$
by rule (ii.a). This is a profitable deviation for each agent $i\in
N^{\alpha}$. Therefore, $f\circ\alpha$ is not Bayesian
implementable. $\quad\quad\square$

\section{A quantum Bayesian mechanism}
Following Ref. \cite{qmd2011}, here we will propose a quantum
Bayesian mechanism to modify the sufficient conditions for Bayesian
implementation. According to Eq (4) in Ref. \cite{Flitney2007},
two-parameter quantum strategies are drawn from the set:
\begin{equation}
\hat{\omega}(\theta,\phi)\equiv \begin{bmatrix}
  e^{i\phi}\cos(\theta/2) & i\sin(\theta/2)\\
  i\sin(\theta/2) & e^{-i\phi}\cos(\theta/2)
\end{bmatrix},
\end{equation}
$\hat{\Omega}\equiv\{\hat{\omega}(\theta,\phi):\theta\in[0,\pi],\phi\in[0,\pi/2]\}$,
$\hat{J}\equiv\cos(\gamma/2)\hat{I}^{\otimes
n}+i\sin(\gamma/2)\hat{\sigma}_{x}^{\otimes
  n}$ (where $\gamma\in[0,\pi/2]$ is an entanglement measure, $\sigma_{x}$ is Pauli matrix),
$\hat{I}\equiv\hat{\omega}(0,0)$,
$\hat{D}_{n}\equiv\hat{\omega}(\pi,\pi/n)$,
$\hat{C}_{n}\equiv\hat{\omega}(0,\pi/n)$.

Without loss of generality, we assume that:\\
1) Each agent $i$ has a quantum coin $i$ (qubit) and a classical
card $i$. The basis vectors $|C\rangle=(1,0)^{T}$,
$|D\rangle=(0,1)^{T}$ of a quantum coin denote head up and tail
up respectively.\\
2) Each agent $i$ independently performs a local unitary operation
on his/her own quantum coin. The set of agent $i$'s operation is
$\hat{\Omega}_{i}=\hat{\Omega}$. A strategic operation chosen by
agent $i$ is denoted as $\hat{\omega}_{i}\in\hat{\Omega}_{i}$. If
$\hat{\omega}_{i}=\hat{I}$, then
$\hat{\omega}_{i}(|C\rangle)=|C\rangle$,
$\hat{\omega}_{i}(|D\rangle)=|D\rangle$; If
$\hat{\omega}_{i}=\hat{D}_{n}$, then
$\hat{\omega}_{i}(|C\rangle)=|D\rangle$,
$\hat{\omega}_{i}(|D\rangle)=|C\rangle$. $\hat{I}$ denotes
``\emph{Not flip}'', $\hat{D}_{n}$ denotes ``\emph{Flip}''.\\
3) The two sides of a card are denoted as Side 0 and Side 1. The
information written on the Side 0 (or Side 1) of card $i$ is denoted
as $card(i,0)$ (or $card(i,1)$). A typical card written by agent $i$
is described as $c_{i}=(card(i,0),card(i,1))$, where
$card(i,0),card(i,1)\in T_{i}\times\mathcal{F} \times
\mathbb{Z}_{+}$. The set of $c_{i}$ is
denoted as $C_{i}$.\\
4) There is a device that can measure the state of $n$ coins and
send messages to the designer.

A \emph{quantum Bayesian mechanism}
$\Gamma^{Q}_{B}=((\hat{\Sigma}_{i})_{i\in N},\hat{g})$ describes a
strategy set $\hat{\Sigma}_{i}=\{\hat{\sigma}_{i}:
T_{i}\mapsto\hat{\Omega}_{i}\times C_{i}\}$ for each agent $i$ and
an outcome function $\hat{g}:\otimes_{i\in
N}\hat{\Omega}_{i}\times\prod_{i\in N}C_{i}\mapsto\mathcal{F}$. A
strategy profile is
$\hat{\sigma}=(\hat{\sigma}_{i},\hat{\sigma}_{-i})$, where
$\hat{\sigma}_{-i}: T_{-i}\mapsto\otimes_{j\neq
i}\hat{\Omega}_{j}\times\prod_{j\neq i}C_{j}$. A \emph{Bayesian Nash
equilibrium} of $\Gamma^{Q}_{B}$ is a strategy profile
$\hat{\sigma}^{*}=(\hat{\sigma}^{*}_{1},\cdots,\hat{\sigma}^{*}_{n})$
such that for every $i\in N$ and for every $t_{i}\in T_{i}$,
\begin{equation*}
U_{i}(\hat{g}(\hat{\sigma}^{*})|t_{i})\geq
U_{i}(\hat{g}(\hat{\sigma}^{*}_{-i},\hat{\sigma}'_{i})|t_{i}), \quad
\forall \hat{\sigma}'_{i}: T_{i}\mapsto\hat{\Omega}_{i}\times C_{i}.
\end{equation*}

\begin{figure}[!t]
\centering
\includegraphics[height=2.7in,clip,keepaspectratio]{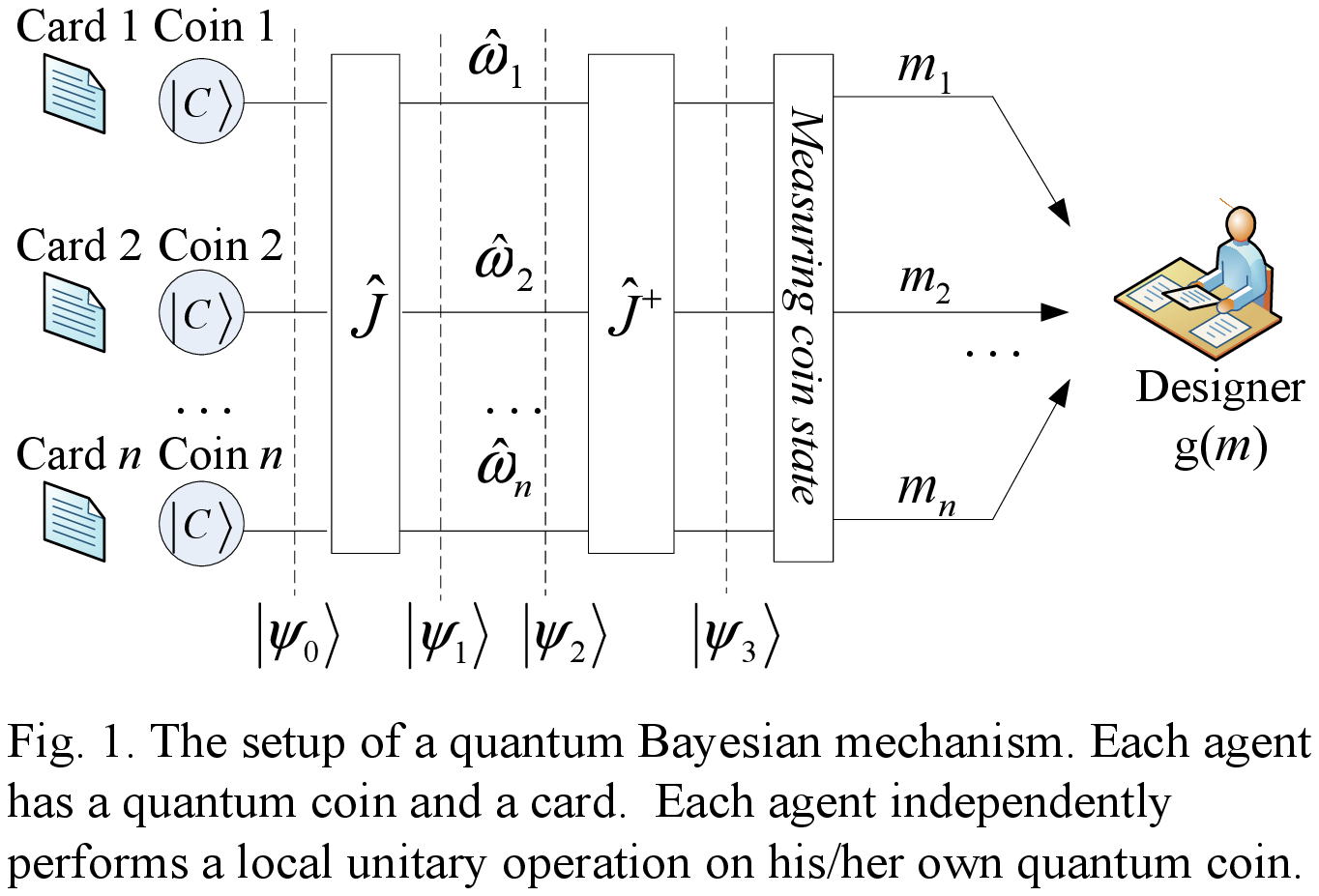}
\end{figure}

The setup of the quantum Bayesian mechanism
$\Gamma^{Q}_{B}=((\hat{\Sigma}_{i})_{i\in N},\hat{g})$ is depicted
in Fig. 1. The working steps of $\Gamma^{Q}_{B}$ are given as
follows:\\
Step 1: Nature selects a state $t\in T$ and assigns $t$ to the
agents. Each agent $i$ knows $t_{i}$ and $q_{i}(t_{-i}|t_{i})$. The
state of each quantum coin is set as $|C\rangle$. The initial state
of the $n$ quantum coins is
$|\psi_{0}\rangle=\underbrace{|C\cdots CC\rangle}\limits_{n}$.\\
Step 2: If $f$ is multi-Bayesian monotonic, then go to Step 4.\\
Step 3: Each agent $i$ sets
$c_{i}=((t_{i},f_{i},z_{i}),(t_{i},f_{i},z_{i}))$,
$\hat{\omega}_{i}=\hat{I}$. Go to Step 7.\\
Step 4: Each agent $i$ sets
$c_{i}=((\alpha_{i}(t_{i}),f,0),(t_{i},f_{i},z_{i}))$ (where
$\alpha$ is specified in the definition of multi-Bayesian
monotonicity). Let $n$ quantum coins be entangled by $\hat{J}$.
$|\psi_{1}\rangle=\hat{J}|\psi_{0}\rangle$.\\
Step 5: Each agent $i$ independently performs a local unitary
operation $\hat{\omega}_{i}$ on his/her own quantum coin.
$|\psi_{2}\rangle=[\hat{\omega}_{1}\otimes\cdots\otimes\hat{\omega}_{n}]|\psi_{1}\rangle$.\\
Step 6: Let $n$ quantum coins be disentangled by $\hat{J}^{+}$.
$|\psi_{3}\rangle=\hat{J}^{+}|\psi_{2}\rangle$.\\
Step 7: The device measures the state of $n$ quantum coins and sends
$card(i,0)$ (or $card(i,1)$) as $m_{i}$ to the designer if the state
of quantum coin $i$ is $|C\rangle$ (or $|D\rangle$).\\
Step 8: The designer receives the overall message
$m=(m_{1},\cdots,m_{n})$ and let the final outcome
$\hat{g}(\hat{\sigma})=g(m)$ using rules (i)-(iii) specified in the
traditional Bayesian mechanism. END.

Given $n\geq 3$ agents and an SCF $f$, suppose $f$ satisfies
multi-Bayesian monotonicity. For each $i\in N^{\alpha}$, let
$card(i,0)=(\alpha_{i}(t_{i}), f, 0)$,
$card(i,1)=(\alpha_{i}(t_{i}), y^{i}, 0)$; for each $i\notin
N^{\alpha}$, let $card(i,0)=(\alpha_{i}(t_{i}), f, 0)$,
$card(i,1)=(t_{i}, f_{i}, z_{i})$  (where $\alpha$, $N^{\alpha}$,
$y^{i}$ are specified in the definition of multi-Bayesian
monotonicity). We define the payoff to the $n$-th agent as follows:
$\$_{C\cdots CC}$ represents the payoff to the $n$-th agent when the
measured state of $n$ quantum coins in Step 7 of $\Gamma^{Q}_{B}$ is
$\underbrace{|C\cdots CC\rangle}\limits_{n}$; $\$_{C\cdots CD}$
represents the payoff to the $n$-th agent when the measured state of
$n$ quantum coins is $|\underbrace{C\cdots
C}\limits_{n-1}D\rangle$). $\$_{D\cdots DD}$ and $\$_{D\cdots DC}$
are defined similarly.

\textbf{Definition 2}: Given an SCF $f$ satisfying multi-Bayesian monotonicity,
define condition $\lambda^{B}$ as follows: \\
1) $\lambda^{B}_{1}$: Consider the payoff to the $n$-th agent,
$\$_{C\cdots CC}>\$_{D\cdots DD}$, i.e., he/she prefers the expected
payoff of a certain outcome (generated by rule (i)) to the expected
payoff of an uncertain outcome (generated by rule (iii)). \\
2) $\lambda^{B}_{2}$:  Consider the payoff to the $n$-th agent,
$\$_{C\cdots CC}>\$_{C\cdots
CD}[1-\sin^{2}\gamma\sin^{2}(\pi/l)]+\$_{D\cdots
DC}\sin^{2}\gamma\sin^{2}(\pi/l)$.

\textbf{Proposition 2}: In economic environments, consider an SCF
$f$ that is incentive compatible and Bayesian monotonic, if $f$ is
multi-Bayesian monotonic and condition $\lambda^{B}$ is satisfied,
then $f\circ\alpha$ is Bayesian implementable by using the quantum
Bayesian mechanism.\\
\textbf{Proof}: Since $f$ is multi-Bayesian monotonic, then there
exist a deception $\alpha$, $f\circ\alpha\napprox f$, and $2\leq
l\leq n$ agents that satisfy Eq (**), i.e., for each agent $i\in
N^{\alpha}$, there exist $t_{i}\in T_{i}$ and an SCF
$y^{i}\in\mathcal{F}$ such that:
\begin{equation*}
U_{i}(y^{i}\circ\alpha|t_{i})>U_{i}(f\circ\alpha|t_{i}), \quad
\mbox{while }U_{i}(y^{i}_{\alpha_{i}(t_{i})}|t'_{i})\leq
U_{i}(f|t'_{i}), \quad\forall t'_{i}\in T_{i}.
\end{equation*}
Hence, the quantum Bayesian mechanism will enter Step 4. Each agent
$i\in N$ sets $c_{i}=((\alpha_{i}(t_{i}),f,0),(t_{i},f_{i},z_{i}))$.
Let $c=(c_{1},\cdots,c_{n})$. Since condition $\lambda^{B}$ is
satisfied, then similar to the proof of Proposition 2 in Ref.
\cite{qmd2011}, if the $n$ agents choose
$\hat{\sigma}^{*}=(\hat{\omega}^{*}, c)$, where
$\hat{\omega}^{*}=(\underbrace{\hat{I},\cdots,\hat{I}}\limits_{n-l},
\underbrace{\hat{C}_{l},\cdots,\hat{C}_{l}}\limits_{l})$, then
$\hat{\sigma}^{*}\in\mathcal{B}(\Gamma^{Q}_{B})$. In Step 7, the
corresponding measured state of $n$ quantum coins is
$\underbrace{|C\cdots CC\rangle}\limits_{n}$. Hence, for each agent
$i\in N$, $m_{i}=(\alpha_{i}(t_{i}),f,0)$. In Step 8,
$\hat{g}(\hat{\sigma}^{*})=f\circ\alpha\napprox f$.\\
Therefore, $f\circ\alpha$ is implemented by $\Gamma^{Q}_{B}$ in
Bayesian Nash equilibrium. $\quad\quad\square$

\section{An algorithmic Bayesian mechanism}
Following Ref. \cite{sim2011}, in this section we will propose an
algorithmic Bayesian mechanism to help agents benefit from the
quantum Bayesian mechanism in the macro world. In the beginning, we
cite matrix representations of quantum states from Ref.
\cite{sim2011}.

\subsection{Matrix representations of quantum states}
In quantum mechanics, a quantum state can be described as a vector.
For a two-level system, there are two basis vectors: $(1,0)^{T}$ and
$(0,1)^{T}$. In the beginning, we define:
\begin{equation}
|C\rangle=\begin{bmatrix}
  1\\
  0
\end{bmatrix},\quad \hat{I}=\begin{bmatrix}
  1 & 0\\
  0 & 1
\end{bmatrix},\quad \hat{\sigma}_{x}=\begin{bmatrix}
  0 & 1\\
  1 & 0
\end{bmatrix},
|\psi_{0}\rangle=\underbrace{|C\cdots
CC\rangle}\limits_{n}=\begin{bmatrix}
  1\\
  0\\
  \cdots\\
  0
\end{bmatrix}_{2^{n}\times1}
\end{equation}
\begin{align}
&\hat{J}=\cos(\gamma/2)\hat{I}^{\otimes
n}+i\sin(\gamma/2)\hat{\sigma}_{x}^{\otimes n}\\
&\quad=\begin{bmatrix}
  \cos(\gamma/2) &  &  &  &  &  & i\sin(\gamma/2)\\
   & \cdots  & &  & & \cdots  & \\
   &  &  & \cos(\gamma/2) & i\sin(\gamma/2) &  & \\
   &  &  & i\sin(\gamma/2) & \cos(\gamma/2) &  & \\
   & \cdots  & &  &  & \cdots & \\
  i\sin(\gamma/2) &  &  &  &  &  & \cos(\gamma/2)
\end{bmatrix}_{2^{n}\times2^{n}}
\end{align}
For $\gamma=\pi/2$,
\begin{align}
\hat{J}_{\pi/2}=\frac{1}{\sqrt{2}}\begin{bmatrix}
  1 &  &  &  &  &  & i\\
   & \cdots  & &  & & \cdots  & \\
   &  &  & 1 & i &  & \\
   &  &  & i & 1 &  & \\
   & \cdots  & &  &  & \cdots & \\
  i &  &  &  &  &  & 1
\end{bmatrix}_{2^{n}\times2^{n}},\quad
\hat{J}^{+}_{\pi/2}=\frac{1}{\sqrt{2}}\begin{bmatrix}
  1 &  &  &  &  &  & -i\\
   & \cdots  & &  & & \cdots  & \\
   &  &  & 1 & -i &  & \\
   &  &  & -i & 1 &  & \\
   & \cdots  & &  &  & \cdots & \\
  -i &  &  &  &  &  & 1
\end{bmatrix}_{2^{n}\times2^{n}}
\end{align}

\subsection{A simulating algorithm}
Similar to Ref. \cite{sim2011}, in the following we will propose a
simulating algorithm that simulates the quantum operations and
measurements in Steps 4-7 of the quantum Bayesian mechanism given in
Section 4. The inputs and outputs of the algorithm are adjusted to
the case of Bayesian implementation. The factor $\gamma$ is also set
as its maximum $\pi/2$. For $n$ agents, the inputs and outputs of
the algorithm are illustrated in Fig. 2. The \emph{Matlab} program
is given in Fig. 3, which is cited from Ref. \cite{sim2011}.

\begin{figure}[!t]
\centering
\includegraphics[height=2.8in,clip,keepaspectratio]{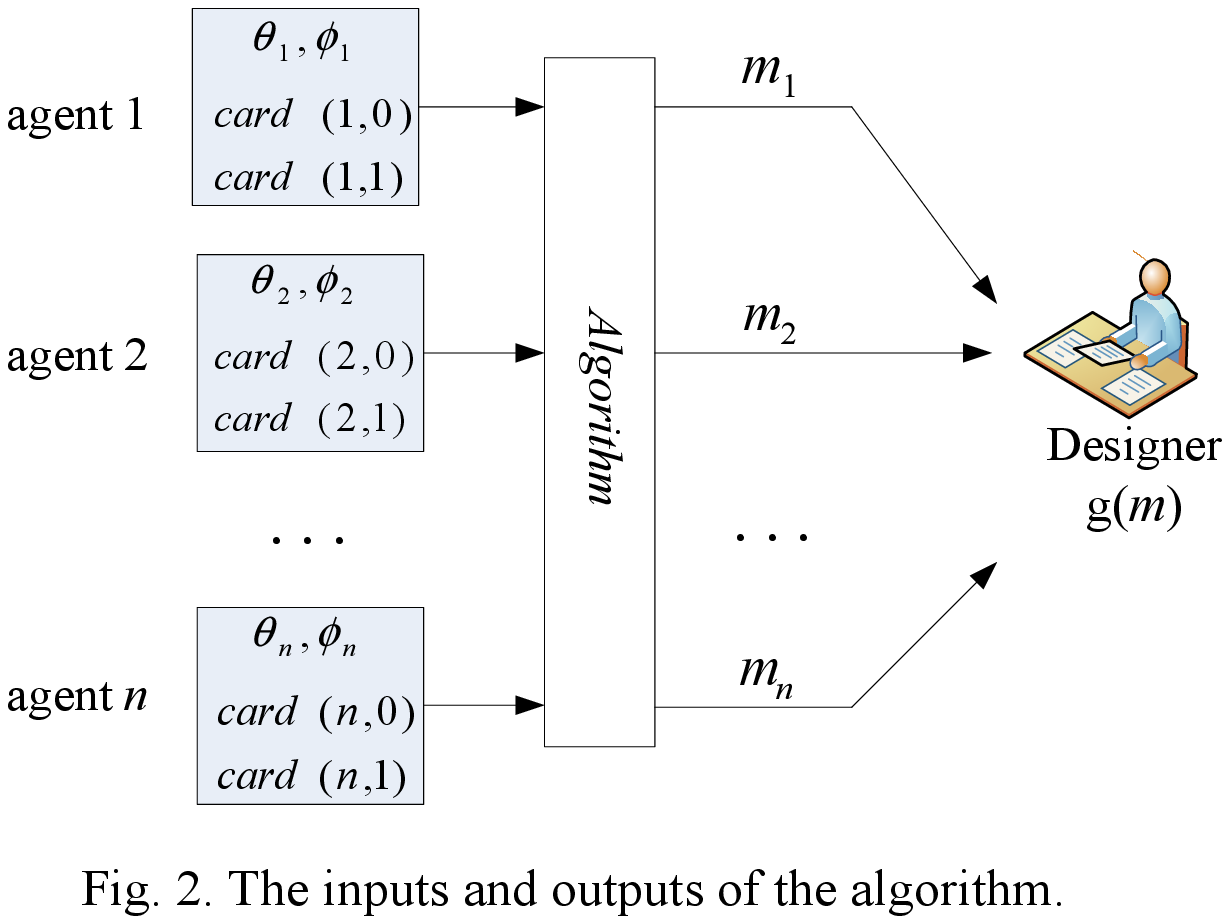}
\end{figure}

\textbf{Inputs}:\\
1) $\theta_{i}$, $\phi_{i}$, $i=1,\cdots,n$: the parameters of agent
$i$'s local operation $\hat{\omega}_{i}$,
$\theta_{i}\in[0,\pi],\phi_{i}\in[0,\pi/2]$.\\
2) $card(i,0), card(i,1)$, $i=1,\cdots,n$: the information written
on the two sides of agent $i$'s card, where $card(i,0),card(i,1)\in
T_{i}\times\mathcal{F} \times \mathbb{Z}_{+}$.

\textbf{Outputs}:\\
$m_{i}$, $i=1,\cdots,n$: the agent $i$'s message that is sent to the
designer, $m_{i}\in T_{i}\times\mathcal{F} \times \mathbb{Z}_{+}$.

\textbf{Procedures of the algorithm}:\\
Step 1: Reading parameters $\theta_{i}$ and $
\phi_{i}$ from each agent $i\in N$ (See Fig. 3(a)).\\
Step 2: Computing the leftmost and rightmost columns of
$\hat{\omega}_{1}\otimes\cdots\otimes\hat{\omega}_{n}$
(See Fig. 3(b)).\\
Step 3: Computing the vector representation of
$|\psi_{2}\rangle=[\hat{\omega}_{1}\otimes\cdots\otimes\hat{\omega}_{n}]
\hat{J}_{\pi/2}|\psi_{0}\rangle$.\\
Step 4: Computing the vector representation of
$|\psi_{3}\rangle=\hat{J}^{+}_{\pi/2}|\psi_{2}\rangle$.\\
Step 5: Computing the probability distribution
$\langle\psi_{3}|\psi_{3}\rangle$ (See Fig. 3(c)).\\
Step 6: Randomly choosing a ``collapsed'' state from the set of all
$2^{n}$ possible states $\{\underbrace{|C\cdots
CC\rangle}\limits_{n}, \cdots,\underbrace{|D\cdots
DD\rangle}\limits_{n}\}$ according to the probability
distribution $\langle\psi_{3}|\psi_{3}\rangle$.\\
Step 7: For each $i\in N$, the algorithm sends $card(i,0)$ (or
$card(i,1)$) as $m_{i}$ to the designer if the $i$-th basis vector
of the ``collapsed'' state is $|C\rangle$ (or $|D\rangle$) (See Fig.
3(d)).

\subsection{An algorithmic version of the quantum Bayesian mechanism}
In the quantum Bayesian mechanism
$\Gamma^{Q}_{B}=((\hat{\Sigma}_{i})_{i\in N},\hat{g})$, the key
parts are quantum operations and measurements, which are restricted
by current experimental technologies. In Section 5.2, these parts
are replaced by a simulating algorithm which can be easily run in a
computer. Now we update the quantum Bayesian mechanism
$\Gamma^{Q}_{B}=((\hat{\Sigma}_{i})_{i\in N},\hat{g})$ to an
\emph{algorithmic Bayesian mechanism}
$\widetilde{\Gamma}_{B}=((\widetilde{\Sigma}_{i})_{i\in
N},\widetilde{g})$, which describes a strategy set
$\widetilde{\Sigma}_{i}=\{\widetilde{\sigma}_{i}: T_{i}\mapsto
[0,\pi]\times[0,\pi/2]\times C_{i}\}$ for each agent $i$ and an
outcome function
$\widetilde{g}:[0,\pi]^{n}\times[0,\pi/2]^{n}\times\prod_{i\in
N}C_{i}\rightarrow\mathcal{F}$, where $n\geq 3$, $C_{i}$ is the set
of agent $i$'s card $c_{i}=(card(i,0), card(i,1))$. A typical
message sent by agent $i$ is denoted by $(\theta_{i}, \phi_{i},
t_{i}, f_{i}, z_{i}, t'_{i}, f'_{i}, z'_{i})$.

A strategy profile is
$\widetilde{\sigma}=(\widetilde{\sigma}_{i},\widetilde{\sigma}_{-i})$,
where $\widetilde{\sigma}_{-i}: T_{-i}\mapsto
[0,\pi]^{n-1}\times[0,\pi/2]^{n-1}\times\prod_{j\neq i}C_{j}$. A
Bayesian Nash equilibrium of $\widetilde{\Gamma}_{B}$ is a strategy
profile
$\widetilde{\sigma}^{*}=(\widetilde{\sigma}^{*}_{1},\cdots,\widetilde{\sigma}^{*}_{n})$
such that for any agent $i\in N$ and for all $t_{i}\in T_{i}$,
\begin{equation*}
U_{i}(\widetilde{g}(\widetilde{\sigma}^{*})|t_{i})\geq
U_{i}(\widetilde{g}(\widetilde{\sigma}^{*}_{-i},\widetilde{\sigma}'_{i})|t_{i}),
\quad \forall \widetilde{\sigma}'_{i}: T_{i}\mapsto
[0,\pi]\times[0,\pi/2]\times C_{i}.
\end{equation*}
Since the factor $\gamma$ is set as its maximum $\pi/2$, the
condition $\lambda^{B}$ in the quantum Bayesian mechanism shall be
updated as $\lambda^{B\pi/2}$. $\lambda^{B\pi/2}_{1}$ is the same as
$\lambda^{B}_{1}$; $\lambda^{B\pi/2}_{2}$ is revised as follows:
Consider the payoff to the $n$-th agent, $\$_{C\cdots
CC}>\$_{C\cdots CD}\cos^{2}(\pi/l)+\$_{D\cdots DC}\sin^{2}(\pi/l)$.

\textbf{Working steps of the algorithmic Bayesian mechanism
$\widetilde{\Gamma}_{B}$}:

Step 1: Given an SCF $f$, if $f$ is multi-Bayesian monotonic,
go to Step 3. \\
Step 2: Each agent $i$ sends $(t_{i},f_{i},z_{i})$ as $m_{i}$ to the
designer. Go to Step 5.\\
Step 3: Each agent $i\in N^{\alpha}$ sets
$card(i,0)=(\alpha_{i}(t_{i}), f, 0)$,
$card(i,1)=(\alpha_{i}(t_{i}), y^{i}, 0)$; each agent $i\notin
N^{\alpha}$ sets $card(i,0)=(\alpha_{i}(t_{i}), f, 0)$,
$card(i,1)=(t_{i}, f_{i}, z_{i})$  (where $\alpha$, $N^{\alpha}$,
$y^{i}$ are specified in the definition of multi-Bayesian
monotonicity). Then each agent $i$ submits $\theta_{i}$,
$\phi_{i}$, $card(i,0)$ and $card(i,1)$ to the simulating algorithm.\\
Step 4: The simulating algorithm runs and outputs
messages $m_{1},\cdots,m_{n}$ to the designer.\\
Step 5: The designer receives the overall message
$m=(m_{1},\cdots,m_{n})$ and let the final outcome be $g(m)$ using
rules (i)-(iii) of the traditional Bayesian mechanism. END.

\subsection{New results for Bayesian implementation}
\textbf{Proposition 3:} In economic environments, given an SCF $f$
that is incentive compatible and Bayesian monotonic:\\
1) If $f$ is multi-Bayesian monotonic and condition
$\lambda^{B\pi/2}$ is satisfied, then $f\circ\alpha$ is Bayesian
implementable by using the algorithmic Bayesian mechanism.\\
2) If $f$ is not multi-Bayesian monotonic, then $f$ is Bayesian implementable.\\
\textbf{Proof}: 1) Since $f$ is multi-Bayesian monotonic,
then $\widetilde{\Gamma}_{B}$ enters Step 3. \\
Each agent $i\in N^{\alpha}$ sets $card(i,0)=(\alpha_{i}(t_{i}), f,
0)$, $card(i,1)=(\alpha_{i}(t_{i}), y^{i}, 0)$; each agent $i\notin
N^{\alpha}$ sets $card(i,0)=(\alpha_{i}(t_{i}), f, 0)$,
$card(i,1)=(t_{i}, f_{i}, z_{i})$  (where $\alpha$, $N^{\alpha}$,
$y^{i}$ are specified in the definition of multi-Bayesian
monotonicity). Then each agent $i$ submits $\theta_{i}$, $\phi_{i}$,
$card(i,0)$ and $card(i,1)$ to the simulating algorithm. Since
condition $\lambda^{B\pi/2}$ is satisfied, then similar to the proof
of Proposition 1 in Ref. \cite{sim2011}, if the $n$ agents choose
$\widetilde{\sigma}^{*}=(\widetilde{\sigma}^{*}_{i})_{i\in N}$,
where for $1\leq i\leq (n-l)$, $\theta_{i}=\phi_{i}=0$; for
$(n-l+1)\leq i\leq n$, $\theta_{i}=0$, $\phi_{i}=\pi/l$, then
$\widetilde{\sigma}^{*}\in\mathcal{B}(\widetilde{\Gamma}_{B}))$.
\\
In Step 6 of the simulating algorithm, the corresponding measured
state is $\underbrace{|C\cdots CC\rangle}\limits_{n}$. Hence, in
Step 7 of the simulating algorithm,
$m_{i}=card(i,0)=(\alpha_{i}(t_{i}),f,0)$ for each agent $i\in N$.
Finally, in Step 5 of $\widetilde{\Gamma}_{B}$,
$\widetilde{g}(\widetilde{\sigma}^{*})=g(m)=f\circ\alpha\napprox f$.\\
Therefore, $f\circ\alpha$ is implemented by $\widetilde{\Gamma}_{B}$
in Bayesian Nash equilibrium.\\
2) If $f$ is not multi-Bayesian monotonic, then
$\widetilde{\Gamma}_{B}$ is reduced to the traditional Bayesian
mechanism. Since the SCF $f$ is incentive compatible and Bayesian
monotonic, then it is Bayesian implementable. $\quad\square$

\section{Conclusions}
This paper follows the series of papers on quantum mechanisms
\cite{qmd2011, sim2011}, and generalizes the quantum and algorithmic
mechanisms in Refs. \cite{qmd2011, sim2011} to Bayesian
implementation. It can be seen that for $n$ agents, the time
complexity of quantum and algorithmic Bayesian mechanisms are $O(n)$
and $O(2^{n})$ respectively. Although current experimental
technologies restrict the quantum Bayesian mechanism to be
commercially available, for small-scale cases (e.g., less than 20
agents \cite{sim2011}), the algorithmic Bayesian mechanism can help
agents benefit from quantum Bayesian mechanism just in the macro
world.

\section*{Acknowledgments}
The author is very grateful to Ms. Fang Chen, Hanyue Wu
(\emph{Apple}), Hanxing Wu (\emph{Lily}) and Hanchen Wu
(\emph{Cindy}) for their great support.

\newpage
\begin{figure}[!t]
\centering
\includegraphics[height=3.9in,clip,keepaspectratio]{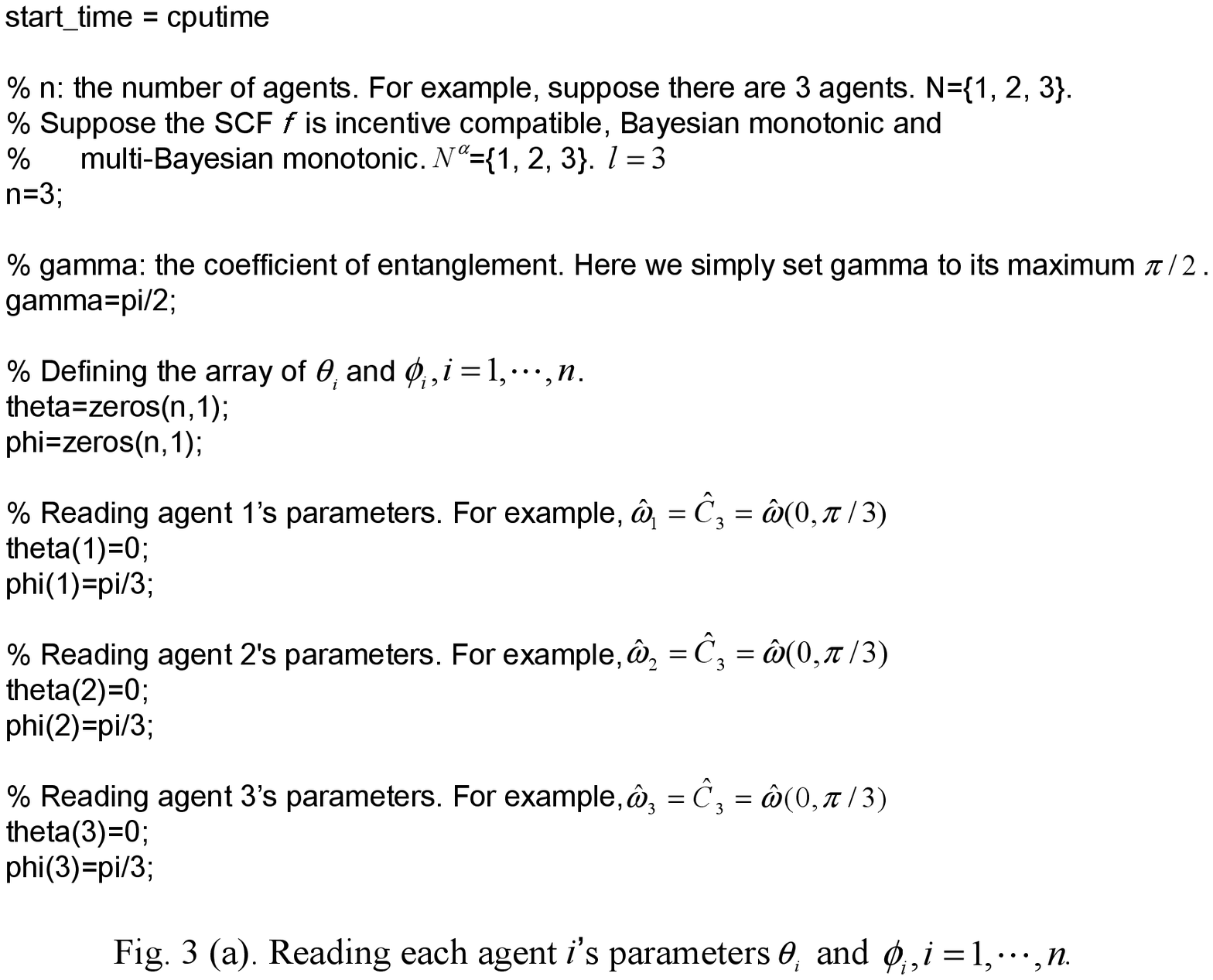}
\end{figure}

\begin{figure}[!t]
\centering
\includegraphics[height=4.7in,clip,keepaspectratio]{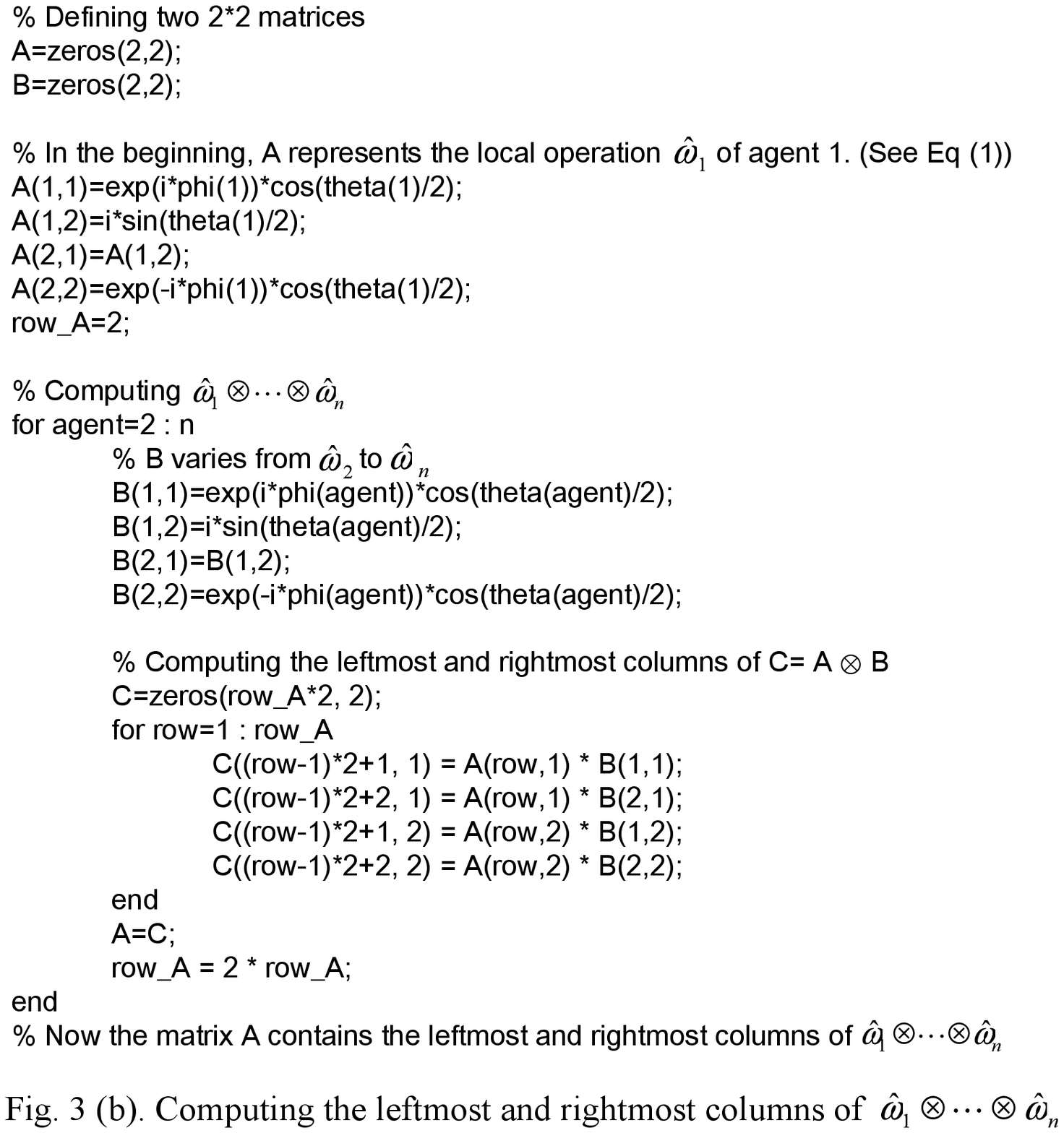}
\end{figure}

\begin{figure}[!t]
\centering
\includegraphics[height=2.6in,clip,keepaspectratio]{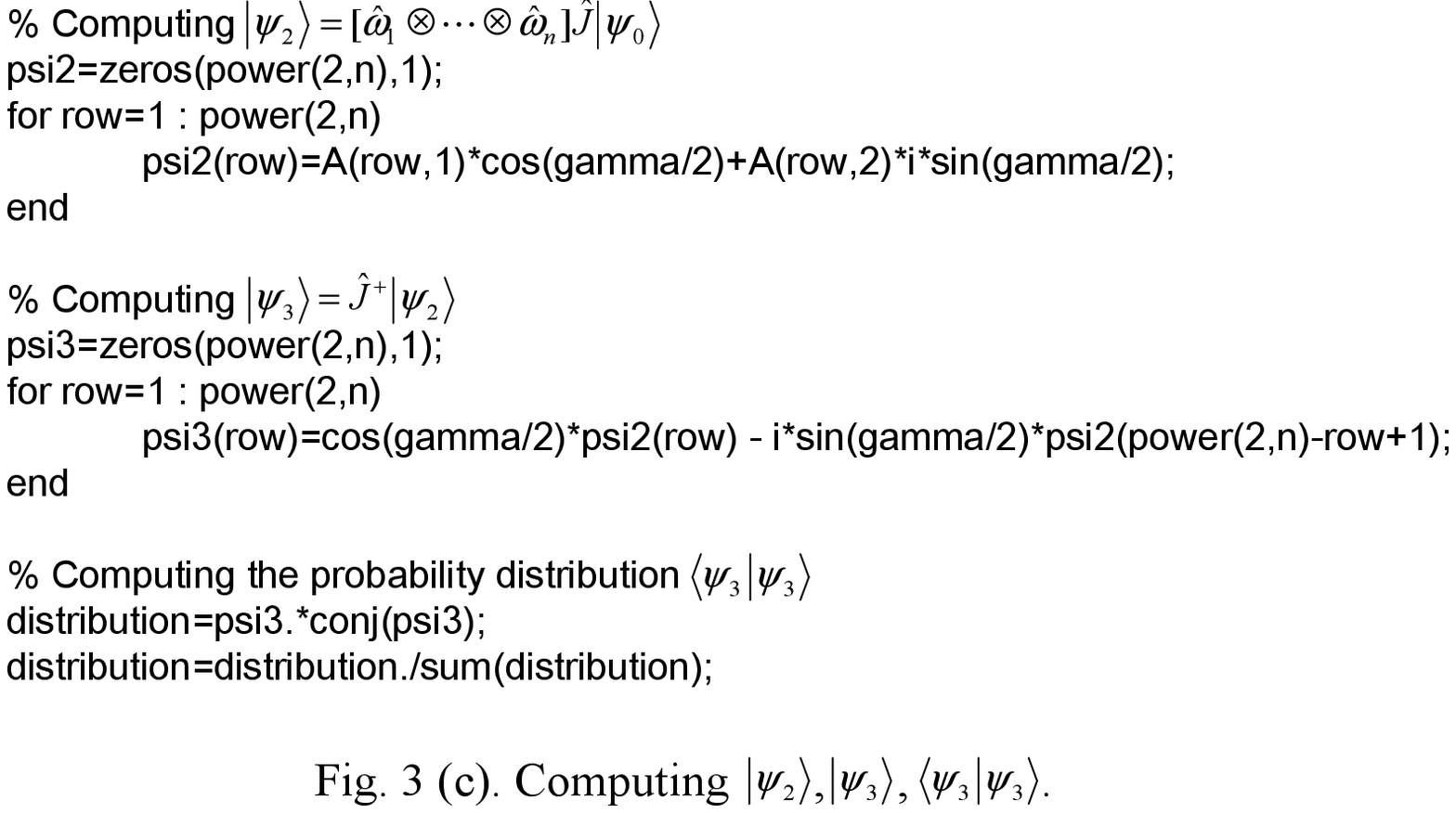}
\end{figure}

\begin{figure}[!t]
\centering
\includegraphics[height=5.8in,clip,keepaspectratio]{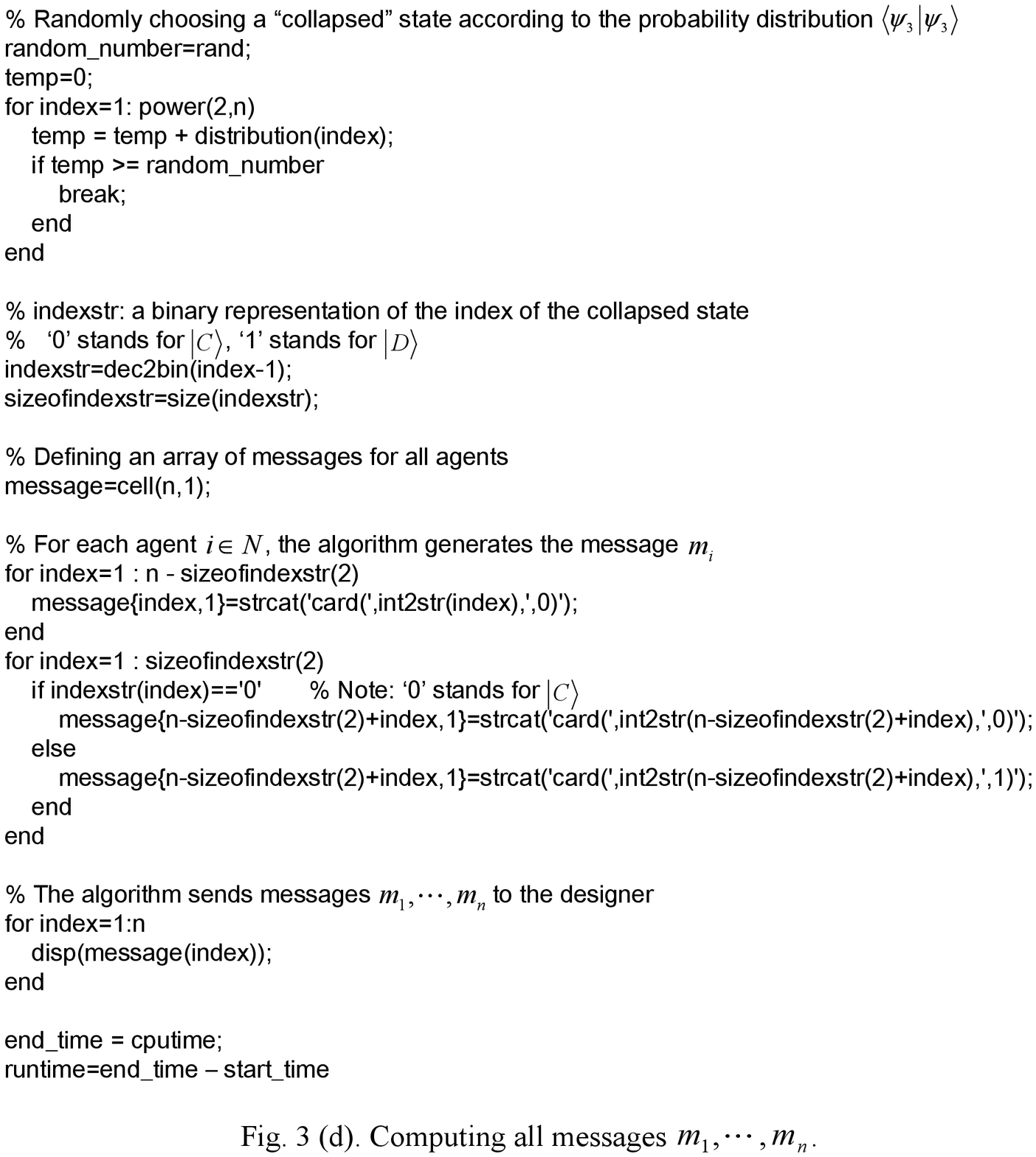}
\end{figure}

\begin{thebibliography}{99}
\bibitem{Maskin1999}
E. Maskin, Nash equilibrium and welfare
optimality, \emph{Rev. Econom. Stud.} \textbf{66} (1999) 23-38.

\bibitem{PS1986}
A. Postlewaite and D. Schmeidler, Implementation in differential
information economies. \emph{Journal of Economic theory},
\textbf{39} (1986) 14-33.

\bibitem{PS1989}
T.R. Palfrey and S. Srivastava, Implementation with incomplete
information in exchange economies, \emph{Econometrica}, \textbf{57}
(1989) 115-134.

\bibitem{Jackson1991}
M.O. Jackson, Bayesian implementation. \emph{Econometrica},
\textbf{59} (1991) 461-477.

\bibitem{qmd2011}
H. Wu, Quantum mechanism helps agents combat ``bad'' social choice
rules. \emph{International Journal of Quantum Information},
\textbf{9} (2011) 615-623.\\
http://arxiv.org/abs/1002.4294

\bibitem{sim2011}
H. Wu, On amending the Maskin's theorem by using complex numbers.
\emph{Games and Economic Behavior}, 2011 (submitted).
http://arxiv.org/abs/1004.5327

\bibitem{Serrano2004}
R. Serrano, The theory of implementation of social choice rules,
\emph{SIAM Review} \textbf{46} (2004) 377-414.

\bibitem{MWG1995}
Mas-Colell, A., MD Whinston, and JR Green, Microeconomic Theory.
Oxford University Press, Oxford, 1995.

\bibitem{Matsushima1993}
H. Matsushima, Bayesian monotonicity with side payments,
\emph{Journal of Economic Theory} \textbf{59} (1993) 107-121.

\bibitem{Flitney2007}
A.P. Flitney and L.C.L. Hollenberg, Nash equilibria in quantum games
with generalized two-parameter strategies, \emph{Phys. Lett. A}
\textbf{363} (2007) 381-388.
\end{thebibliography}
\end{document}